\documentclass[aps,prd]{revtex4}
\usepackage{amssymb,amsmath,amsfonts,comment,float,latexsym,graphicx,epstopdf}
\usepackage{color}
\usepackage{fancyvrb}
\usepackage{chngcntr}
\usepackage{etoolbox}

\newcommand{\be}{\begin{equation}}
\newcommand{\ee}{\end{equation}}
\newcommand{\ba}{\begin{eqnarray}}
\newcommand{\ea}{\end{eqnarray}}

\newcommand{\grts}{\raise.3ex\hbox{$>$\kern-.75em\lower1ex\hbox{$\sim$}}}
\newcommand{\lets}{\raise.3ex\hbox{$<$\kern-.75em\lower1ex\hbox{$\sim$}}}

\newcommand{\f}{\frac}
\newcommand{\lf}{\left}
\newcommand{\rh}{\right}
\patchcmd{\section}
  {\centering}
  {\raggedright}
  {}
  {}
\patchcmd{\subsection}
  {\centering}
  {\raggedright}
  {}
  {}

\begin{document}

\title{The Lee-Wick Extension of the Two-Higgs Doublet Model}

\author{Aria R. Johansen}\email[]{aria.r.johansen@gmail.com}
\author{Marc Sher}\email[]{mtsher@wm.edu}
\author{Keith Thrasher}\email[]{rkthrasher@email.wm.edu}

\affiliation{High Energy Theory Group, College of William and Mary, Williamsburg, Virginia 23187, U.S.A.}
\date{\today}

\begin{abstract}
\begin{center}
\vspace{-2mm}
\large{Abstract}
\end{center}
\vspace{-3mm}
The Lee-Wick Standard Model is a highly constrained model which solves the gauge hierarchy problem at the expense of including states with negative norm.   It appears to be macroscopically causal and consistent.    This model is extended by considering the two-Higgs doublet extension of the Lee-Wick model.  Rewriting the Lagrangian using auxiliary fields introduces two additional doublets of Lee-Wick partners.   The model is highly constrained, with only one or two additional parameters beyond that of the usual two-Higgs doublet model, and yet there are four doublets.   Mass relations are established by diagonalizing the mass matrices and further constraints are established by studying results from $B\rightarrow \tau \nu$, neutral $B$-meson mixing, and $B \rightarrow X_s \gamma$. The prospects of detecting evidence for this model at the LHC are discussed.
\end{abstract}

\maketitle

\section{Introduction}

Fifty years ago, T.D. Lee and G.C. Wick \cite{Lee:1969fy, Lee:1970iw} proposed a model in an attempt to soften the ultraviolet divergences of QED.  This model added a quartic kinetic energy term to the Lagrangian.  The resulting propagator has two poles, resulting in two physical states, the effects of which cancel quadratic divergences.  Using an auxiliary field method, one can show that the effective Lagrangian consists of only operators of dimension less than or equal to four, with one of the fields having a negative kinetic energy term, leading to apparent violations of causality.  Lee and Wick showed, along with Cutkosky et al.\cite{Cutkosky:1969fq} and Coleman \cite{Coleman:1969xz}, that while microcausality is violated, unitarity is preserved and at the macroscopic level there are no logical paradoxes.

Motivated by the cancelation of divergences, Grinstein, O'Connell and Wise \cite{Grinstein:2007mp} constructed the Lee-Wick Standard Model (LWSM).  As in the original Lee-Wick model, all particle states come with Lee-Wick partners which have negative kinetic terms.  These Lee-Wick partners cancel the quadratic divergences in the scalar propagator, thus solving the hierarchy problem in a manner similar to supersymmetry. Grinstein, et al \cite {Grinstein:2007iz} also demonstrated that the scattering of longitudinally polarized massive vector bosons satisfied perturbative unitarity.  Explicitly, they later showed that unitarity and Lorentz-invariance are preserved in the S-matrix to all orders and that causality arises as an emergent macroscopic phenomenon\cite{ Grinstein:2008bg}.

Since the Grinstein et al papers, there have been numerous phenomenological studies of the LWSM, including, but not limited to the study the possibility of observing the microcausality violation at colliders \cite{ Alvarez:2009zz,Alvarez:2009af,Alvarez:2011ah,Figy:2011yu,Lebed:2012zv,Chivukula:2010kx}, the effects of the LWSM on precision electroweak measurements \cite{Underwood:2008cr,Alvarez:2008za,Carone:2009nu,Chivukula:2010nw,Lebed:2012ab,Carone:2014kla}, and finite temperature effects \cite{Fornal:2009xc,Bhattacharya:2011bb,Lebed:2013gua}.  The LW partners of the light quarks and gluons must be relatively heavy, $O(10)$ TeV, in order to avoid detection.  However, the LW spectrum, as in the case of the Minimal Super Symmetric Model(MSSM), is not degenerate.  Thus one can have some states relatively heavy while others, canceling quadratic divergences, can be lighter\cite{Carone:2008bs}.  Just as in the MSSM, one would expect the LW partners to the electroweak gauge bosons, the Higgs, top quark, and left-handed bottom quark to be in the effective low-energy theory in order to avoid substantial fine-tuning of the hierarchy. The focus here is on the Higgs sector.

The model consists of a Two Higgs Doublet with only one additional parameter beyond the Standard Model.  As a result, all additional scalar masses, the ratio of vacuum expectation values and mixing angles are determined by this parameter.  The strongest bound on this parameter comes from B physics\cite{Carone:2009nu}, and gives typical scalar masses lower bounds of approximately 450 GeV.

Given an N-Higgs doublet model, the Lee-Wick extension will be a 2N-Higgs doublet model.   This article explores the simplest extension of the Higgs sector, the Two Higgs Doublet Model (2HDM), with the simplest LW extension resulting in a Four Higgs Doublet Model, with only one additional parameter beyond the 2HDM.  The new model, with only one additional parameter but eight additional Higgs fields and their numerous couplings and mixings, will then be very tightly constrained.  The parameters of the 2HDM, in models with no tree-level flavor-changing neutral currents, can be expressed in terms of the scalar masses and mixings.  In addition to the type-I 2HDM, the charged Higgs can be light, close enough in mass to the top quark, and it will be interesting to see if that can be maintained.

In the next section, the LWSM is presented, following earlier works.  Section III contains the Lee-Wick 2HDM  (LW2HDM), where the various constraints are presented.  The constraints from low-energy physics (primarily B physics) are in Section IV, and the results at current and prospects at future colliders are discussed in Section V.  Mass matrices and coupling constant relations are given in the Appendix.
 
\section{The Lee-Wick Standard Model Higgs Sector}

The Higgs sector of the Lee-Wick Standard Model (LWSM) is given by a Lagrangian with a higher derivative kinetic term \cite{Grinstein:2007mp}
\be \label{eq:laghd}
{\cal L}_{HD} = (D_\mu \hat H)^{\dagger} (D^\mu \hat H) 
-\frac{1}{m^2_{\tilde h}} (D_\mu D^\mu \hat H)^{\dagger} (D_\nu D^\nu \hat H) - V(\hat H).
\ee
The potential takes the usual form  
\be
V(\hat H) = \frac{\lambda}{4} \left(\hat H^\dagger \hat H -\f{v^2}{2}\right)^2.
\ee
To eliminate the higher-derivative term, an auxiliary field $\tilde{H}$ is introduced, giving the Lagrangian
\be\label{eq:laf}
{\cal L}_{AF} = (D_\mu \hat H)^{\dagger} (D^\mu \hat H) + (D_\mu \hat H)^\dagger (D^\mu \tilde H) 
+ (D_\mu \tilde H)^\dagger (D^\mu \hat H) + m_{\tilde h}^2\tilde H^\dagger \tilde H  - V(\hat H).
\ee
The higher derivative Lagrangian is reproduced by substituting the equation of motion for the auxiliary field.   The kinetic terms are diagonlized by redefining $\hat{H} = H - \tilde{H}$:
\be \label{eq:GaugeHiggs} 
{\cal L} = (D_\mu H)^{\dagger} (D^\mu H) 
- (D_\mu \tilde H)^{\dagger} (D^\mu \tilde H) + m_{\tilde h}^2\tilde H^\dagger\tilde H- V (H - \tilde H).
\ee
The higher derivative term has been eliminated by introducing the LW field $\tilde{H}$ which has the opposite sign kinetic term of the usual particle.

A gauge is chosen so that 
\be
H = \begin{pmatrix}
0 \\ \f{v+h}{\sqrt{2}}
\end{pmatrix}, \;\;\;
\tilde H = \begin{pmatrix}
\tilde h^{+} \\ \f{\tilde h + i \tilde P}{\sqrt{2}}
\end{pmatrix}.
\ee
where $v \approx 246$ GeV, the Higgs vev.

The neutral scalar mass matrix must now be diagonalized.   It is of the form
\be\label{massmatrixform}
{\cal L}_M =  -\frac{1}{2}\begin{pmatrix} m^2_h & -m^2_h \\ -m^2_h & -(m^2_{\tilde{h}}-m^2_h)\end{pmatrix}
\ee

Normally, when one chooses to diagonalize a scalar mass matrix, an orthogonal representation is used since that will not affect the structure of the kinetic terms.   However, in this case, one of the kinetic terms has a negative coefficient, and an orthogonal transformation will not preserve this form.   Instead, a symplectic transformation must be used.
\begin{equation} \label{symprot}
\left( \begin{array}{c} h \\ \tilde h \end{array} \right) = \left(
\begin{array}{cc} \cosh \eta  & \sinh \eta \\ \sinh \eta & \cosh
\eta \end{array} \right) \left( \begin{array}{c} h_0 \\ \tilde h_0
\end{array} \right) \, ,
\end{equation}
where subscript $0$ indicates a mass eigenstate.  The mixing angle $\eta$ satisfies
\begin{equation} \label{tanhform}
\tanh 2\eta = \frac{-2m_h^2/m_{\tilde h}^2}{1-2m_h^2/m_{\tilde h}^2}
\,\,\,\,\,\mbox{or}\,\,\,\,\,
 \tanh \eta= -\frac{m_{h_0}^2}{m_{\tilde h_0}^2} 
\end{equation}
with mass eigenvalues
\be \label{eq:eigenmass}
m_{h_0}^2 = \frac{m_{\tilde h}^2}{2} \left( 1 -
\sqrt{ 1 -\frac{4 m_h^2}{m_{\tilde h}^2} } \right) \,\,\,\,\, \mbox{ and } \,\,\,\,\,
m_{\tilde h_0}^2 = \frac{m_{\tilde h}^2}{2} \left( 1 +
\sqrt{ 1 -\frac{4 m_h^2}{m_{\tilde h}^2} } \right) \,.
\ee
It is easy to see that the LW pseudoscalar $P$ and the LW charged scalar $\tilde{h}^{\pm}$ have the same mass and that the heavier of the neutral scalars has the negative kinetic energy term.    The masses of the neutral scalars are related to the mass of the charged scalar by
\be \label{eq:massrelation} 
m_{h_0}^2 + m_{\tilde h_0}^2 = m_{\tilde h}^2.
\ee

The ratio of the couplings of the neutral Higgs bosons to their value in the Standard Model, $g_{XY}$, are \cite{Carone:2014kla}
\begin{equation}
g_{h_0 t \overline{t}} = g_{h_0 b \overline{b}}=g_{h_0 \tau \tau} = e^{-\eta} \,\,\, ,
\end{equation}
\begin{equation}
g_{h_0 WW} = g_{h_0 ZZ} = \cosh\eta \,\,\,\, ,
\end{equation}
\begin{equation}
g_{\tilde{h}_0 t \overline{t}} = g_{\tilde{h}_0 b \overline{b}}=g_{\tilde{h}_0 \tau \tau} = - e^{-\eta} \,\,\, ,
\end{equation}
\begin{equation}
g_{\tilde{h}_0 WW} = g_{\tilde{h}_0 ZZ} = \sinh\eta \,\,\,\, .
\end{equation}
An important property of these couplings is that the coupling of the light Higgs to the SM gauge bosons is greater than those in the SM.   In most extensions of the SM, the couplings are suppressed, but this is an exception.

Note that this model is similar to a  type-II 2HDM, with $\tan\beta = 1$ and some minus signs in the vertices and propagators.   As a result, a single parameter, the Lee-Wick scale, gives all mixing angles, Yukawa couplings, masses and interactions of the LW Higgs bosons.   This makes the model very predictive.    In Ref.\cite{Carone:2009nu} and \cite{Carone:2014kla}, bounds on the model from B-meson and Z decays and LHC studies of the light Higgs boson are examined.   The strongest of these constraints comes from radiative B-decays and gives a lower bound on the heavy neutral (charged) scalar of $445$ ($463$) GeV.

The LW2HDM can be expected to have the same number of parameters as the standard 2HDMs, with the addition of the Lee-Wick scale.  Given the larger number of states in this model, it will also be highly predictive.

\section{The LW Two-Higgs Doublet Model}

It is straightforward to generalize the LW higher derivative Lagrangian from the previous section.
\be
{\cal{L}}_{HD}= (D_\mu \hat{H}_1)^\dagger(D^\mu \hat{H}_1) - \frac{1}{m^2_{\tilde{h}_1}}(D_\mu D^\mu \hat{H}_1)(D_\nu D^\nu \hat{H}_1)+(D_\mu \hat{H}_2)^\dagger(D^\mu \hat{H}_2) - \frac{1}{m^2_{\tilde{h}_2}}(D_\mu D^\mu \hat{H}_2)(D_\nu D^\nu \hat{H}_2) - V(\hat{H}_1,\hat{H}_2)
\ee
Here, $V(\hat{H}_1,\hat{H}_2)$ is the standard Two-Higgs Doublet Model potential (see Ref. \cite{Branco:2011iw}), where $H_1$ and $H_2$ are the Two-Higgs Doublets.  The potential contains:
\be
\begin{aligned}
V(\hat{H}_1,\hat{H}_2) &= m_{11}^2  \hat{H}_1^{\dagger} \hat{H}_1+m_{22}^2 \hat{H}_1^{\dagger} \hat{H}_1 -m_{12}^2( \hat{H}_1^{\dagger} \hat{H}_2+ \hat{H}_2^{\dagger} \hat{H}_1) +\frac{1}{2}\lambda_1 \left( \hat{H}_1^{\dagger} \hat{H}_1 \right)^2+\frac{1}{2}\lambda_2 \left( \hat{H}_2^{\dagger} \hat{H}_2 \right)^2\\
&+\lambda_3\hat{H}_1^{\dagger} \hat{H}_1\hat{H}_2^{\dagger} \hat{H}_2
+ \lambda_4\hat{H}_1^{\dagger} \hat{H}_2\hat{H}_2^{\dagger} \hat{H}_1 
+\frac{1}{2} \lambda_5 \left( \left( \hat{H}_1^{\dagger} \hat{H}_2 \right)^2+\left( \hat{H}_2^{\dagger} \hat{H}_1\right)^2 \right).
\end{aligned}
\ee
where the $\lambda_i$ terms are then the coupling constants between the Higgs fields

Note that there are two different Lee-Wick scales in this Lagrangian.   As will be seen, the mass matrices can easily be diagonalized if these scales are equal.  This assumption will be made here, and the possible consequences of relaxing the assumption will be discussed later.

Following the same procedure as before, by introducing auxiliary fields, then redefining the fields in order to diagonalize the kinetic energy terms, the new Lagrangian is
\be 
\begin{aligned}
{\cal L} &= (D_\mu H_1)^\dagger (D^\mu  H_1) - (D_\mu \tilde{H}_1)^\dagger (D^\mu  \tilde{H}_1) + (D_\mu H_2)^\dagger (D^\mu  H_2) - (D_\mu \tilde{H}_2)^\dagger (D^\mu  \tilde{H}_2)+ \\
& \,\,\,\,\,\,\,\,\, m_{\tilde{h}}^2(\tilde{H}_1^\dagger\tilde{H}_1+ \tilde{H}_2^\dagger\tilde{H}_2)- V(H_1-\tilde{H_1},H_2-\tilde{H}_2)
\end{aligned}
\ee

Minimizing the potential, then evaluating the second derivatives with respect to each field gives the mass matrices for this model.  The generated matrices are in the Appendix.  As expected, there are four neutral scalars, four pseudoscalars and four charged scalars.  The charged and pseudoscalars have a zero diagonal element when they are diagonalized, corresponding to the Goldstone bosons.   These diagonal elements are {\bf not} necessarily eigenvalues obtained from solving the secular determinant, since a symplectic transformation does not preserve the form of the kinetic terms.

To diagonalize the mass matrices, an orthogonal transformation is applied to the upper $2\times2$ and lower $2\times2$ blocks.  For the charged and pseudoscalar mass matrices, these transformations are both just a rotation by $\beta$ (as in the usual Two Higgs Doublet Model).  For the neutral scalar mass matrix, the rotation is defined as $\alpha$.  Upon performing these transformations, the charged Higgs mass matrix is
\be
\left(
\begin{array}{cccc}
 0 & 0 & 0 & 0 \\
 0 & -\frac{\left(v_1^2+v_2^2\right) \left(v_1 v_2 \left(\lambda _4+\lambda _5\right)-2 m_{12}^2\right)}{2 v_1 v_2} & 0 & \frac{\left(v_1^2+v_2^2\right) \left(v_1 v_2 \left(\lambda _4+\lambda _5\right)-2 m_{12}^2\right)}{2 v_1 v_2} \\
 0 & 0 & -m_{\tilde{h}}^2 & 0 \\
 0 & \frac{\left(v_1^2+v_2^2\right) \left(v_1 v_2 \left(\lambda _4+\lambda _5\right)-2 m_{12}^2\right)}{2 v_1 v_2} & 0 & \frac{2 m_{12}^2 \left(v_1^2+v_2^2\right)-v_1 v_2 \left(2 m_{\tilde{h}}^2+\left(v_1^2+v_2^2\right) \left(\lambda _4+\lambda _5\right)\right)}{2 v_1 v_2} \\
\end{array}
\right)
\ee

Note the zero (indicating the presence of the Goldstone boson) on the diagonal.   One mass is the Lee-Wick scale (resulting from the negative kinetic term, and positive mass-squared term).   The remaining $2 \times 2$ submatrix is precisely of the form as Eq. \ref{massmatrixform}, and thus can be diagonalized with a symplectic transformation, resulting in

\be
diag(0,\,m_{H^\pm_0}^2 ,\,-m_{\tilde{H}^{\prime\pm}_0}^2,\,-m_{\tilde{H}^\pm_0}^2 )=
\left(\begin{array}{cccc}
 0 & 0 & 0 & 0 \\
 0 & -\frac{1}{2} m_{\tilde{h}}^2 \left(A-1\right) & 0 & 0 \\
 0 & 0 & -m_{\tilde{h}}^2 & 0 \\
 0 & 0 & 0 & -\frac{1}{2} m_{\tilde{h}}^2 \left(A+1\right) \\
\end{array}\right),
\ee
where $A = \sqrt{\frac{m_{\tilde{h}}^2+2 \left(v^2 \left(\lambda _4+\lambda _5\right)-2 M_{12}^2\right)}{m_{\tilde{h}}^2}}$.
The three masses clearly obey the relation $m_{H^\pm_0}^2 - m_{\tilde{H}^{\pm}_0}^2 = m_{\tilde{H}^{\prime \pm}_0}^2$. The pseudoscalar masses have precisely the same relationship. The scalars obey a similar relationship, with masses $m_{h_0}^2 - m_{\tilde{h}_0}^2  = m_{H_0}^2 - m_{\tilde{H}_0}^2$ which are given in the Appendix.  These relations are absolute predictions of the model.

The symplectic transformation in each case, similar to the LWSM case,  are given by $\tanh\Psi = -m_{0}^2/\tilde{m}_0^2$, where $m_0$ and $\tilde{m}_0$ are the physical masses. In the case of the charged Higgs, for example, the mixing angle of the symplectic transformation that diagonalizes the mass matrix is given by $\tanh \theta = m_{H^{\prime \pm}_0}^2/m_{\tilde{H}^{\prime \pm}_0}^2$ .  For the pseudoscalar case, a similar result is found.   For the neutral scalar case, there are two symplectic rotations needed to diagonalize the mass matrix.  The neutral scalar masses and scalar couplings can be found in terms of the masses and mixing angles in the Appendix.

In the Two-Higgs Doublet model, the observed scalar at $125$ GeV has couplings to the $W^\pm$ and $Z$ which are $\sin(\beta-\alpha)$ times that of the SM.  The dual scalar, $H$, has couplings which are $\cos(\alpha-\beta)$ times that of the SM.   The pseudoscalar and charged scalar have no tree-level couplings to gauge bosons.   Similarly, in this model the couplings to the gauge bosons are

\be
h_0ZZ=h_0W^{+}W^{-}=\cosh \left(\psi _1\right) \sin(\beta-\alpha)
\ee

\be
\tilde{h}_0ZZ=h_0W^{+}W^{-}=\sinh \left(\psi _1\right) \sin(\beta-\alpha)
\ee

\be
H_0ZZ=h_0W^{+}W^{-}=\cosh \left(\psi _2\right) \cos (\alpha -\beta )
\ee

\be
\tilde{H}_0ZZ=h_0W^{+}W^{-}=\sinh \left(\psi _2\right) \cos (\alpha -\beta )
\ee
where $\psi_1,\psi_2$ are the symplectic transformation angles for the neutral scalars.

%%%%%%%%%%%%%%%%%%%%%%%%%%%%%

The determination of the neutral scalars coupling to the weak gauge bosons allows for the Yukawa couplings to be resolved. In the 2HDM, the Yukawa couplings are dependent upon the type of 2HDM being studied. The Higgs doublets take the form

\be
\Phi_j= \left(
\begin{array}{c}
\phi^+_j \\
\frac{v_j+\rho_j+i \eta_j}{\sqrt{2}}
\end{array}
\right).
\ee

In the type-I 2HDM, $\Phi_2$ couples to both $u^i_R$ and  $d^i_R$, while in the type-II model $\Phi_2$ couples to $u^i_R$ and $\Phi_1$ couples to $d^i_R$. Considering the LW extensions of these two models, the Yukawa interactions take the form

% Yukawa interaction Lagrangian%
\begin{equation}
\begin{aligned}
\mathcal{L}^{LW 2 HDM}_{Yukawa} &\supset - \sum_{f=u,d} \frac{m_f}{v} 
 \left( \sum_{\substack{H=h_0,\tilde{h}_0, \\ H_0,\tilde{H}_0}} \xi^f _H \bar{f}fH 
-i  \sum_{\substack{A=A_0,\\ \tilde{A_0},\tilde{A^\prime_0}}} \xi^f _A \bar{f}fA  \right) \\
&-\frac{\sqrt{2}}{v}\sum_{\substack{H^+=H^+_0, \\ \tilde{H}^+_0,\tilde{H}^{\prime +}_0}}
\left[ V_{ud}\bar{u}\left( m_u \xi^u_{H^+} P_L + m_d \xi^d_{H^+} P_R \right)d H^{+} 
 +H.C.\right]
 \end{aligned}
\end{equation}
where the expressions for the parameters $\xi^f$ are found in Table \ref{tab:couplings}. The Yukawa couplings of the neutral scalar Higgs and associated LW neutral scalar Higgs to the quarks only differs by a sign. This same feature exists in the LWSM.  In general, the sign difference is also present for the the pseudo-scalar and charged Higgs. When the LW scale goes to infinity, one recovers usual 2HDM couplings.

% Table of Coupling Coefficients %
\begingroup\makeatletter\def\f@size{6}\check@mathfonts
\begin{table}[h]
\begin{tabular}{|c|c|c|}
\hline 
 & \hspace{20mm} Type I \hspace{20mm} &\hspace{20mm} Type II \hspace{20mm} \tabularnewline
\hline 
\hline 
$\xi_{h_{0}}^{u}$ & $e^{-\psi_{1}}\cos(\alpha)\csc(\beta)$ & $e^{-\psi_{1}}\cos(\alpha)\csc(\beta)$\tabularnewline
\hline 
$\xi_{h_{0}}^{d}$ & $e^{-\psi_{1}}\cos(\alpha)\csc(\beta)$ & $-e^{-\psi_{1}}\cos(\alpha)\sec(\beta)$\tabularnewline
\hline 
$\xi_{\tilde{h}_{0}}^{u}$ & $-e^{-\psi_{1}}\cos(\alpha)\csc(\beta)$ & $-e^{-\psi_{1}}\cos(\alpha)\csc(\beta)$\tabularnewline
\hline 
$\xi_{\tilde{h}_{0}}^{d}$ & $-e^{-\psi_{1}}\cos(\alpha)\csc(\beta)$ & $e^{-\psi_{1}}\cos(\alpha)\sec(\beta)$\tabularnewline
\hline 
$\xi_{H_{0}}^{u}$ & $e^{-\psi_{2}}\sin(\alpha)\csc(\beta)$ & $e^{-\psi_{2}}\sin(\alpha)\csc(\beta)$\tabularnewline
\hline 
$\xi_{H_{0}}^{d}$ & $e^{-\psi_{2}}\sin(\alpha)\csc(\beta)$ & $e^{-\psi_{2}}\sin(\alpha)\sec(\beta)$\tabularnewline
\hline 
$\xi_{\tilde{H}_{0}}^{u}$ & $-e^{-\psi_{2}}\sin(\alpha)\csc(\beta)$ & $-e^{-\psi_{2}}\sin(\alpha)\csc(\beta)$\tabularnewline
\hline 
$\xi_{\tilde{H}_{0}}^{d}$ & $-e^{-\psi_{2}}\sin(\alpha)\csc(\beta)$ & $-e^{-\psi_{2}}\sin(\alpha)\sec(\beta)$\tabularnewline
\hline 
$\xi_{A_{0}}^{u}$ & $e^{-\phi}\cot(\beta)$ & $e^{-\phi}\cot(\beta)$\tabularnewline
\hline 
$\xi_{A_{0}}^{d}$ & $-e^{-\phi}\cot(\beta)$ & $e^{-\phi}\tan(\beta)$\tabularnewline
\hline 
$\xi_{\tilde{A}_{0}}^{u}$ & $-e^{-\phi}\cot(\beta)$ & $-e^{-\phi}\cot(\beta)$\tabularnewline
\hline 
$\xi_{\tilde{A}_{0}}^{d}$ & $e^{-\phi}\cot(\beta)$ & $-e^{-\phi}\tan(\beta)$\tabularnewline
\hline 
$\xi_{\tilde{A_{0}^{\prime}}}^{u}$ & -1 & -1\tabularnewline
\hline 
$\xi_{\tilde{A_{0}^{\prime}}}^{d}$ & 1 & 1\tabularnewline
\hline 
$\xi_{H_{0}^{\pm}}^{u}$ & $e^{-\theta}\cot(\beta)$ & $e^{-\theta}\cot(\beta)$\tabularnewline
\hline 
$\xi_{H^{\pm}_{0}}^{d}$ & $-e^{-\theta}\cot(\beta)$ & $e^{-\theta}\tan(\beta)$\tabularnewline
\hline 
$\xi_{\tilde{H_{0}^{\pm}}}^{u}$ & $-e^{-\theta}\cot(\beta)$ & $-e^{-\theta}\cot(\beta)$\tabularnewline
\hline 
$\xi_{\tilde{H_{0}^{\pm}}}^{d}$ & $e^{-\theta}\cot(\beta)$ & $-e^{-\theta}\tan(\beta)$\tabularnewline
\hline 
$\xi_{\tilde{H_{0}^{\prime\pm}}}^{u}$ & -1 & -1\tabularnewline
\hline 
$\xi_{\tilde{H_{0}^{\prime\pm}}}^{d}$ & 1 & 1\tabularnewline
\hline 
\end{tabular}
\caption{Yukawa couplings of the quarks to the Higgs bosons. Angles $\psi_1$ and $\psi_2$ are the symplectic rotations needed to diagonalize the two neutral scalar mass matrix, $\phi$ is the rotation angle to diagonalize the pseudoscalar mass matrix and $\theta$ is the angle which diagonalizes the charged scalar mass matrix. These angles are all determined in terms of the physical particle masses, as described in the text.}
\label{tab:couplings}
\end{table}
\endgroup

For simplicity, it was assumed that the Lee-Wick scales in Eq. (15) were equal. We know of no principle or symmetry that would lead to this equality, although one would not expect qualitative differences. Suppose this assumption is relaxed.    Consider the charged Higgs mass matrix.    Applying orthogonal transformations to the upper and lower $2\times 2$ blocks gives the mass matrix
\be
\left(
\begin{array}{cccc}
 0 & 0 & 0 & 0 \\
 0 & M_{12}^2-\frac{1}{2} \left(\lambda _4+\lambda _5\right) v^2 & 0 & -(M_{12}^2-\frac{1}{2} \left(\lambda _4+\lambda _5\right) v^2) \\
 0 & 0 & -\cos ^2(\beta ) m_{\tilde{h}_1}^2-\sin ^2(\beta ) m_{\tilde{h}_2}^2 & \cos (\beta ) \sin (\beta ) \left(m_{\tilde{h}_1}^2-m_{\tilde{h}_2}^2\right) \\
 0 & -(M_{12}^2-\frac{1}{2} \left(\lambda _4+\lambda _5\right) v^2) & \cos (\beta ) \sin (\beta ) \left(m_{\tilde{h}_1}^2-m_{\tilde{h}_2}^2\right) & \,\,\,\,M_{12}^2-\frac{1}{2} \left(\lambda _4+\lambda _5\right) v^2 -\sin ^2(\beta ) m_{\tilde{h}_1}^2-\cos ^2(\beta ) m_{\tilde{h}_2}^2\\
\end{array}
\right).
\ee
One sees that in the limit in which the scales are equal, this reduces to the previous result.     There is no simple hyperbolic rotation that diagonalizes this mass matrix.    However, one can first consider the case in which the Lee-Wick scales are close together, so that the 3-4 and 4-3 elements of the mass matrix are much smaller than the other terms.     In that case, one can find the masses explicitly and they are given by
(with, as before, the charged Higgs mass-squared being denoted $m^2_{H^\pm_0}$)   $m^2_{H^\pm_0}$, $m^2_{\tilde{h}_1}\cos^2\beta + m^2_{\tilde{h}_2}\sin^2\beta$ and $m^2_{H^\pm_0} + m^2_{\tilde{h}_1}\sin^2\beta + m^2_{\tilde{h}_2}\cos^2\beta$.  

Of course, long before these particles are discovered, it is likely that $\tan\beta$ will have been determined, and thus the Lee-Wick charged scalar masses will determine the two Lee-Wick scales.   However, once those scales are determined, the masses and mixings of the neutral LW scalars and pseudoscalars are completely determined.   This is not a surprise, since we have added an extra parameter, and thus the masses of the charged scalars no longer have the simple relationship from before.   However, the model remains highly predictive, since all of the other LW scalar masses and their mixing angles are then determined.   Note also that these facts are expected to be true even when the mass splitting is not small, although then there is no simple analytic expression for these masses and mixings.

%%%%%%%%%Constraints Section%%%%%%%%%%%%
\section{Low Energy Constraints}

In the analysis of the LWSM, Carone et al. \cite{Carone:2009nu} showed that constraints from B-physics give the strongest bounds on the model.  With the above Yukawa couplings, the constraints can similarly be calculated.  In this section, constraints from  $B^+\longrightarrow \tau^+ \nu_\tau$, $ B_d\bar{B}_d $ mixing, and $ B\longrightarrow X_s \gamma$ are explored, leading to lower bounds for the mass of the charged Higgs, $m_{H^\pm_0}$, and its Lee-Wick partners. 

%%%%%%%%%%%%%%%%%%%%%%%%
\subsection{$B^+\longrightarrow \tau^+ \nu_\tau$ } 
For large $\tan\beta$, the strongest bounds come from the branching ratio of $B^+\longrightarrow \tau^+ \nu_\tau$. In the 2HDM the rate is
\be
\frac{\mathcal{B}\left(B^+\longrightarrow \tau^+ \nu_\tau \right)}{\mathcal{B}\left(B^+\longrightarrow \tau^+ \nu_\tau \right)_{SM}}
=\left(1-\frac{m_B^2 C_i}{m^2_{H^\pm_0}} \right)^2
\ee
where $C_1=\cot^2 \beta$ is the coefficient from the type-I 2HDM, and $C_2=\tan^2 \beta$ is the coefficient from the type-II 2HDM. There are now two additional charged Higgs in the model, making the 2HDM result have an additional two Feynman diagrams resulting in,

\be
\frac{\mathcal{B}\left(B^+\longrightarrow \tau^+ \nu_\tau \right)}{\mathcal{B}\left(B^+\longrightarrow \tau^+ \nu_\tau \right)_{SM}}
=\left(1-\frac{m_B^2 e^{-2\theta} C_i  }{m^2_{H^\pm_0}}
+\frac{m_B^2 e^{-2\theta} C_i  }{m^2_{\tilde{H}^\pm_0}}
+\frac{m_B^2  }{m^2_{\tilde{H}^{\prime \pm}_0}} \right)^2.
\ee

Note the difference in sign in the latter two terms on the left hand side of the above equation.  This is a result of the opposite sign in the propagators of the LW particles.  Taking the limit of the LW scale, $m_{\tilde{h}} \longrightarrow \infty$, recovers the 2HDM result. Plots of the branching ratio for $B^+ \longrightarrow \tau^+ \nu_\tau$ for the type-II model are below in Figure \ref{fig:tauplots}.

\begin{figure}[H]
 \centering
 	\includegraphics[scale = .6]{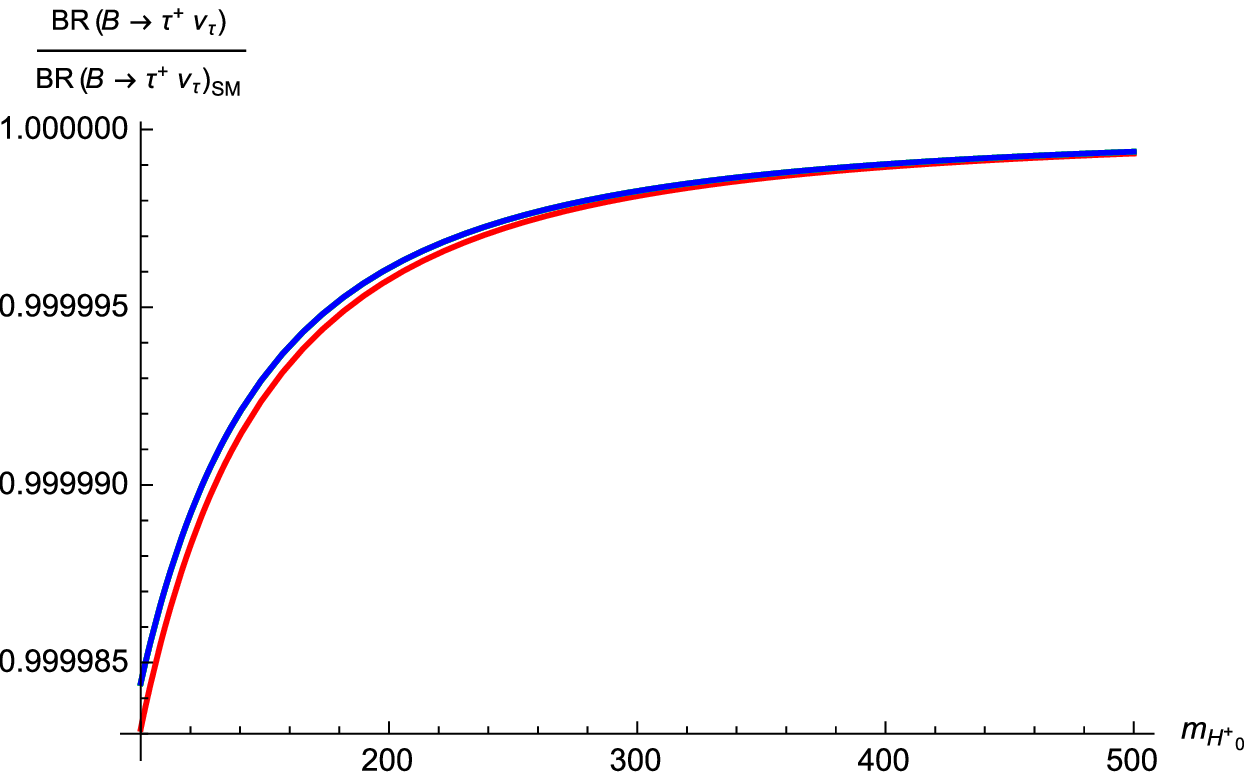} 
	\includegraphics[scale = .6]{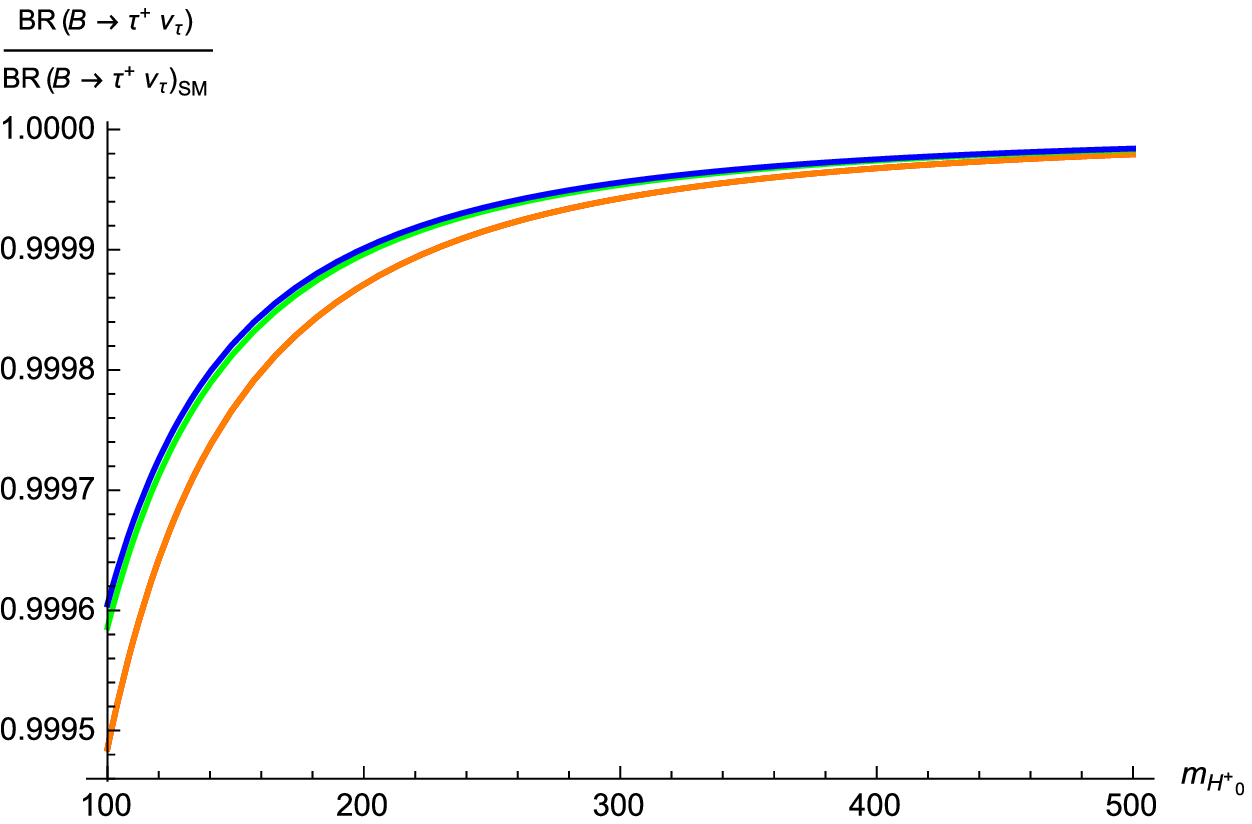}
	\includegraphics[scale = 1]{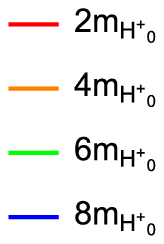} 
\caption{Branching ratio, $\mathcal{B}\left(B^+\longrightarrow \tau^+ \nu \right)$, in the type-II LW2HDM normalized with the standard model result for various LW scales. Left plot shows result for $\tan \beta = 2$ and the right plot  for $\tan \beta = 5$. }  
\label{fig:tauplots}
\end{figure}

The Heavy Flavour Averaging Group\cite{HFAG} combined the results from the experiments BELLE\cite{Hara:2010dk, Ikado:2006un} and BABAR\cite{babar:2010} to find the $\mathcal{B}\left(B^+\longrightarrow \tau^+ \nu \right)$ branching ratio to be $(1.64\pm 0.34)\times 10^{-4}$.  Dividing the HFAG experimental result by the Standard Model predicted result \cite{Czarnecki:1998tn} gives $1.37 \pm 0.39$.  This lower bound on the mass of the charged Higgs in the type-II LW2HDM was established at the 95\% confidence level and is shown in the summary plot at the end of this section, Figure \ref{fig:bounds}.

%%%%%%%%%%%%%%%%%%%%%%%%
\subsection{$B_d\bar{B}_d$ mixing}

In the 2HDM, the result for the mass splitting between $B$ and $\bar{B}$ is identical for both type-I and II 2HDMs. It has been shown that the mass splitting at LO in QCD is \cite{Gunion:1990}

\be
\Delta m_{B_{2HDM}} = \f{G_F^2}{6\pi^2}m_W^2 \lf|V_{tq}V_{tb}^*\rh|^2\ f_B^2 \hat{B}_{B_q} m_B 
\lf( I_{WW}+\cot^2\beta\ I_{WH}+\cot^4\beta \  I_{HH} \rh),
\ee
where  $I_{WW}$ is the contribution from a 2 $W^\pm$ exchange, $I_{WH}$ is the contribution from a single charged Higgs exchange, and $I_{HH}$ is the contribution from a 2 charged Higgs exchange.  Explicitly,

\ba
I_{WW}&=&\f{x}{4} \left(1+\f{9}{\lf(1-x\rh)}-\f{6}{\lf(1-x\rh)^2} 
- \f{6}{x} \lf(\f{x}{1-x}\rh)^3\ln x\right) \, ,\nonumber \\
I_{WH}&=&\f{xy}{4} \left[ -\f{8-2x}{(1-x)(1-y)}+\f{6x\ln x}{(1-x)^2(y-x)} 
+\f{\lf(2x-8y\rh)\ln y}{(1-y)^2(y-x)} \right] \, ,\nonumber \\
I_{HH}&=&\f{xy}{4} \lf[ \f{\lf(1+y\rh)}{\lf(1-y\rh)^2} + \f{2y\ln y}{\lf(1-y\rh)^3} \rh],
\ea
where $x = {m_t^2}/{m_W^2}$ and $y={m_t^2}/{m_{H^+}^2}$ . Making the following modifications allows one to accommodate the additional Higgs into the calculation of $\Delta m_{B}$.

\be
\cot^2 \beta\ I_{WH} \longrightarrow e^{-2\theta} \cot^2 \beta\ I_{WH}(y\rightarrow y_0)-e^{-2\theta} \cot^2 \beta\ I_{WH}(y\rightarrow \tilde{y}_0) -  I_{WH}(y\rightarrow \tilde{y}^\prime_0)
\ee

\be
\cot^4 \beta\ I_{HH} \longrightarrow e^{-4\theta} \cot^4 \beta\ I_{HH}(y\rightarrow y_0)+e^{-4\theta} \cot^4 \beta\ I_{HH}(y\rightarrow \tilde{y}_0) +  I_{HH}(y\rightarrow \tilde{y}^\prime_0)
\ee
where $y_0={m_t^2}/{m_{ H^+_0}^2}$, $\tilde{y}_0={m_t^2}/{m_{ \tilde{H}^+_0}^2}$, and $\tilde{y}^\prime_0={m_t^2}/{m_{\tilde{ H}^{\prime +}_0}^2}$.

From here, the only terms not accounted for are those from mixed charged Higgs exchanges. Making an approximation allows for solving of the mixed charged Higgs exchanges.  Averaging the masses gives
$$
m_{H^+_{12}}=\frac{m_{ H^+_0}+m_{ \tilde{H}^+_0}}{2} \,\,\,\,\,\,\,\,\,
m_{H^+_{13}}=\frac{m_{ H^+_0}+m_{ \tilde{H}^{\prime +}_0}}{2} \,\,\,\,\,\,\,\,
m_{H^+_{23}}=\frac{m_{ \tilde{H}^+_0}+m_{ \tilde{H}^{\prime +}_0}}{2}. 
$$ 
and three additional $I_{HH}$ terms are added where the intermediate Higgs are treated as the averaged masses of the two Higgs being exchanged.The added terms take the form
\vspace{1 mm}
\be
-e^{-4\theta} \cot^4 \beta\ I_{HH}(y\rightarrow y_{12})-e^{-2\theta} \cot^2 \beta\ I_{HH}(y\rightarrow y_{13})+e^{-2\theta} \cot^2 \beta\ I_{HH}(y\rightarrow y_{23}),
\ee
where $y_{ij}=\frac{m_t^2}{m^2_{H^+_{ij}}}$. If the values for the averaged masses are varied between the two masses being averaged, the change in $\Delta m_{B_d}$ falls within the bounds of the uncertainty.  The same modifications were applied to the NLO amplitudes in Ref. \cite{Urban:1997gw}.  

The theoretical uncertainty in $\Delta m_{B_d}$, is primarily dominated by the QCD bag-factor  $f_B^2 \hat{B}_{B_q}$, and is approximated by $\sigma = 0.14\, \Delta m_{B_d}$. 
A $\chi^2$ test,
$$
\chi^2_i=\frac{\left( \mathcal{O}_i^{th}-\mathcal{O}_i^{exp} \right)^2}{\sigma^2_i}
$$
 was used to obtain bounds on the charged Higgs mass, $m_{H^+_0}$, at the 95\% confidence level, corresponding to $\chi^2=3.84$. An experimental value of $\Delta m_{B_d}=(3.337 \pm 0.033) \times 10^{-10}$ MeV \cite{Amsler} was used. Plots of $\Delta m_{B_d}$ at NLO in QCD are given in Figure \ref{fig:dmb}. Values used in the numerical calculation are in the Appendix.  Plots of the excluded region for the charged Higgs mass are shown at the end of the section in Figure \ref{fig:bounds}.

\begin{figure}[H]
 \centering
 	\includegraphics[scale = .6]{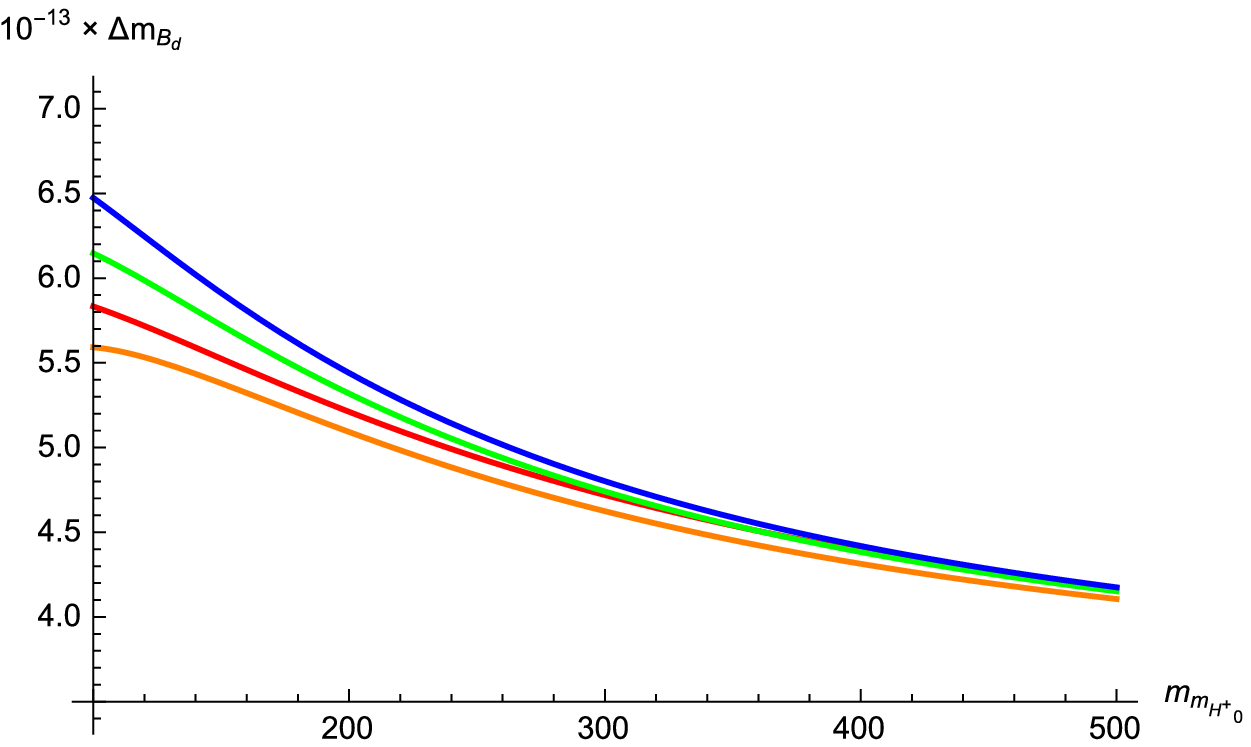} 
	\includegraphics[scale = .6]{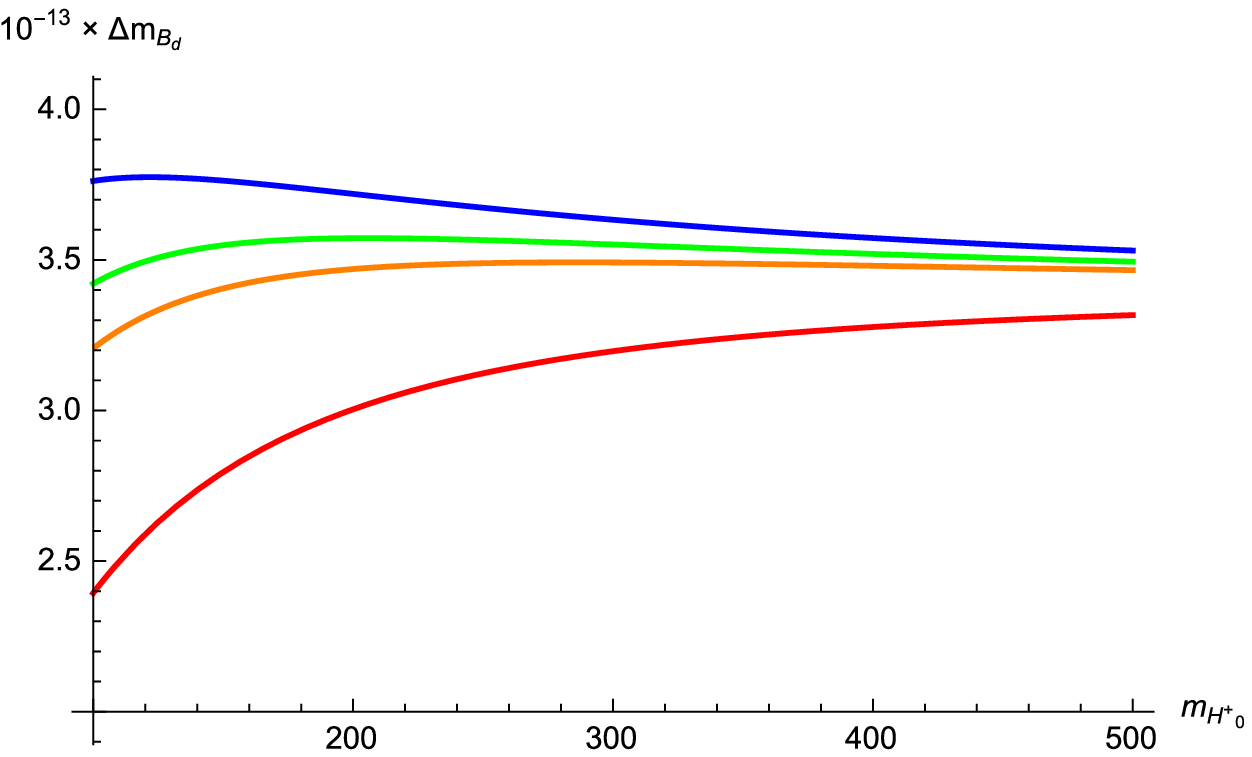} 
	\includegraphics[scale = 1]{Legend.eps} \\
\caption{Plots of $\Delta m_{B_d}$ in $GeV$ given for various LW scales for $\tan\beta=1$ on left and $\tan\beta=2$ on right. Note that the plots all converge to the standard model result in the limit of large $m_{H^\pm_0}$}
\label{fig:dmb}
\end{figure}

%%%%%%%%%%%%%%%%%%%%%%%%
\subsection{ $B\longrightarrow X_s \gamma$ } 

Now considering $B\longrightarrow X_s \gamma$, the LO contribution of the $B\longrightarrow X_s \gamma$ decay is \cite{Grinstein:1987pu}
\be
\mathcal{B}(B\rightarrow X_s\gamma) = 
\mathcal B(B \rightarrow X_ce\bar\nu_e) \left|\f{V_{ts}^*V_{tb}}{V_{cb}}\right|^2 
\f{6\alpha_{em}}{\pi f(m^2_c/m^2_b)} \left|C^0_{7,SM}+C^0_{7,NP}\right|^2,
\ee
where $C^0_7$ are Wilson coefficients. In the type II 2HDM, these coefficients are given by 
\be
C^0_{7,SM}=\frac{x}{24}\left[\frac{-8x^3+3x^2+12x-7+(18x^2-12x)\ln(x)}{(x-1)^4}\right] \, , 
\ee
\be
C^0_{7,NP}=\frac{1}{3}\cot^2(\beta)\; C^0_{7,SM}(x\rightarrow y) 
+ \frac{1}{12}y\left[\frac{-5y^2+y-3+(6y-4)\ln(y)}{(y-1)^3}\right] \, ,
\ee
where $x=\frac{m_t^2}{m_W^2}$ and $y=\frac{m_t^2}{m^2_{H^{+}_0}}$.  For the LW extension of the type-II 2HDM, this becomes
\be
\begin{aligned}
C^0_{7,NP} &=\frac{1}{3}e^{-2\theta} \cot^2(\beta) C^0_{7,SM}(x\rightarrow y) - \frac{1}{12}y\left[\frac{-5y^2+y-3+(6y-4)\ln(y)}{(y-1)^3}\right]  \\
&-\frac{1}{3}e^{-2\theta} \cot^2(\beta) C^0_{7,SM}(x\rightarrow w) - \frac{1}{12}w\left[\frac{-5w^2+w-3+(6w-4)\ln(w)}{(w-1)^3}\right] \\
&-\frac{1}{3} C^0_{7,SM}(x\rightarrow z) - \frac{1}{12}z\left[\frac{-5z^2+z-3+(6z-4)\ln(z)}{(z-1)^3}\right] ,
\end{aligned}
\ee
where $w=\frac{m_t^2}{m^2_{\tilde{H}^+_0}}$ and $z=\frac{m_t^2}{m^2_{\tilde{H}^{\prime +}_0}}$. The function $f(\xi)$, a phase space suppression factor from the semileptonic decay rate, is 
\be
f(\xi)=1-8\xi+8\xi^3-\xi^4-12\xi^2 \ln( \xi).
\ee

\begin{figure}[H]
 \centering
 	\hspace{- 21mm}
 	\includegraphics[scale = .6]{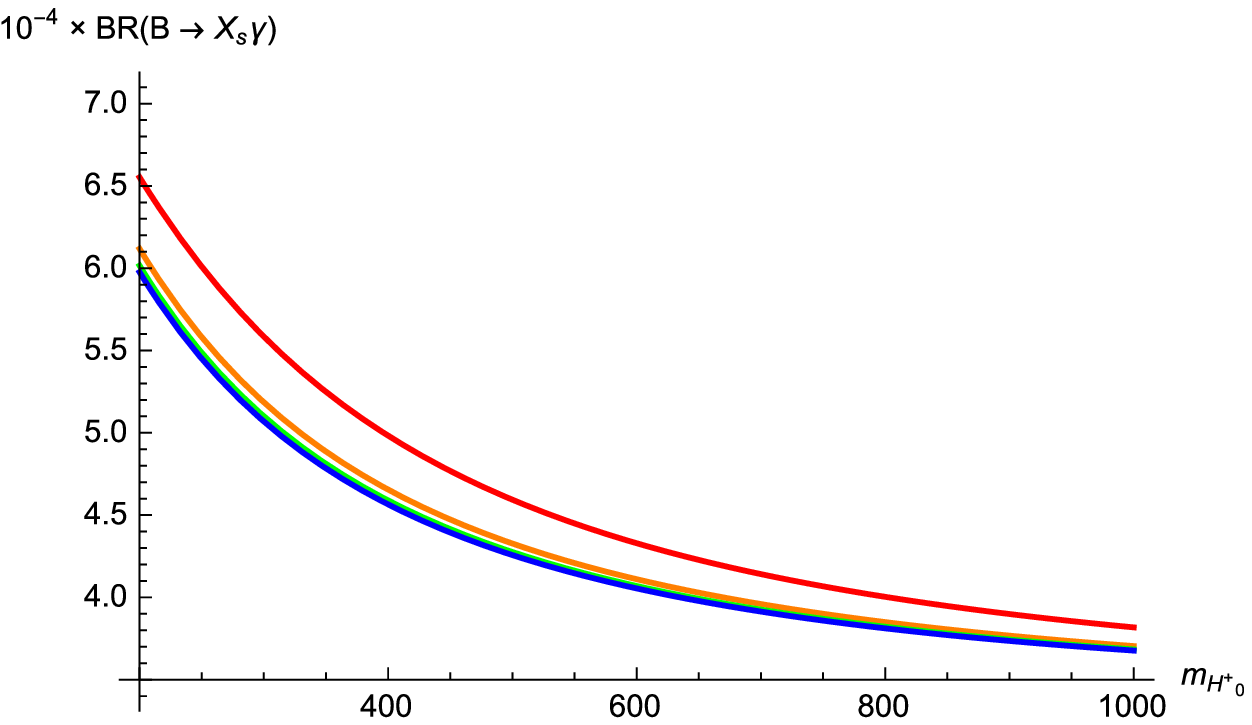} 
	\includegraphics[scale = .6]{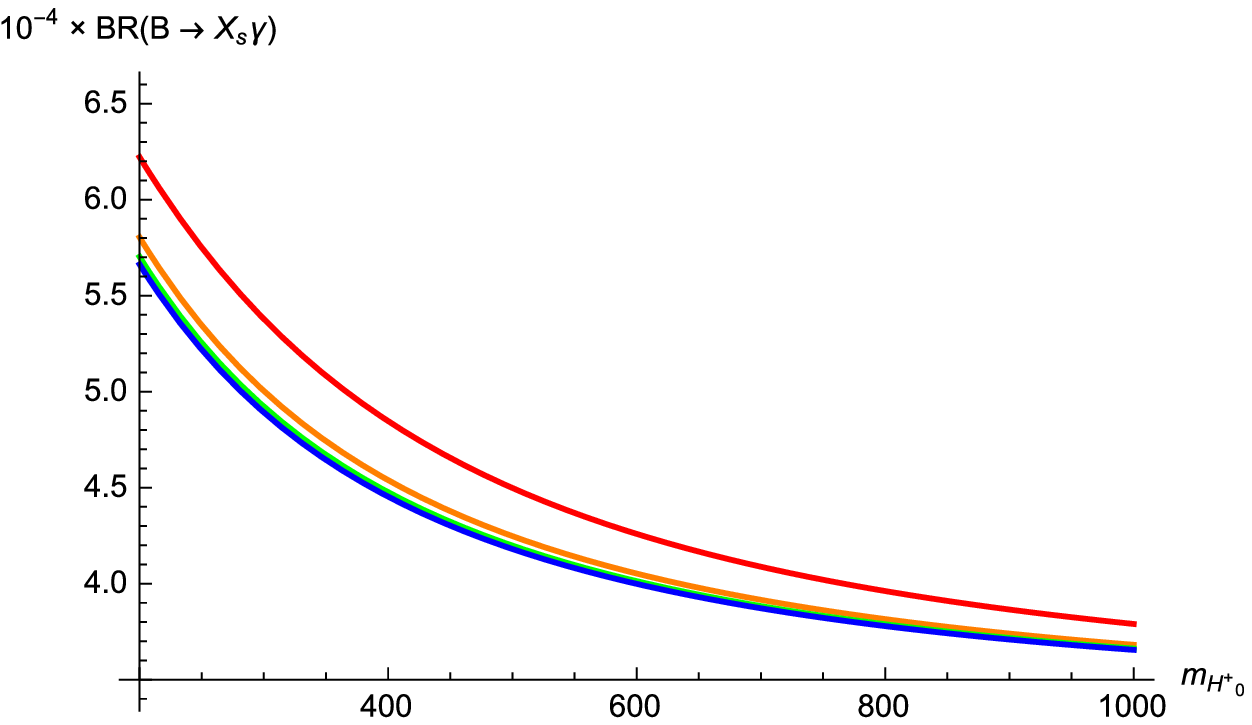} 
	\hspace{0mm}
	\includegraphics[scale = .6]{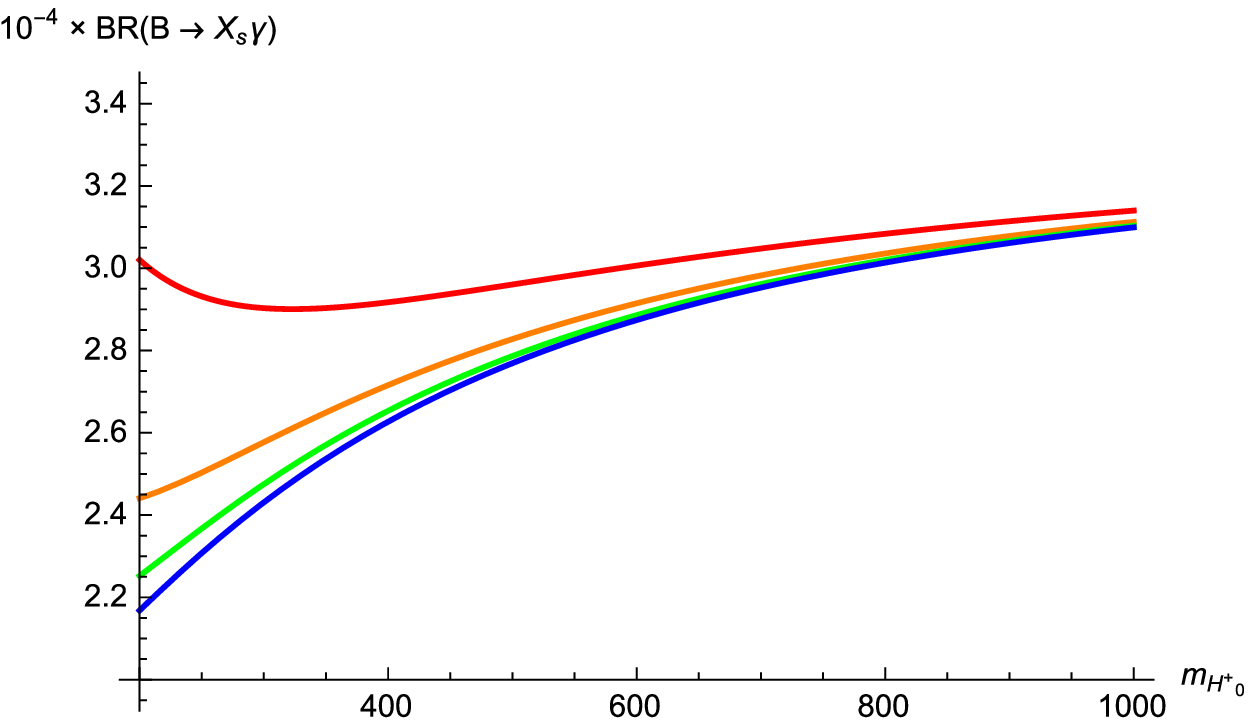} 
	\includegraphics[scale = .6]{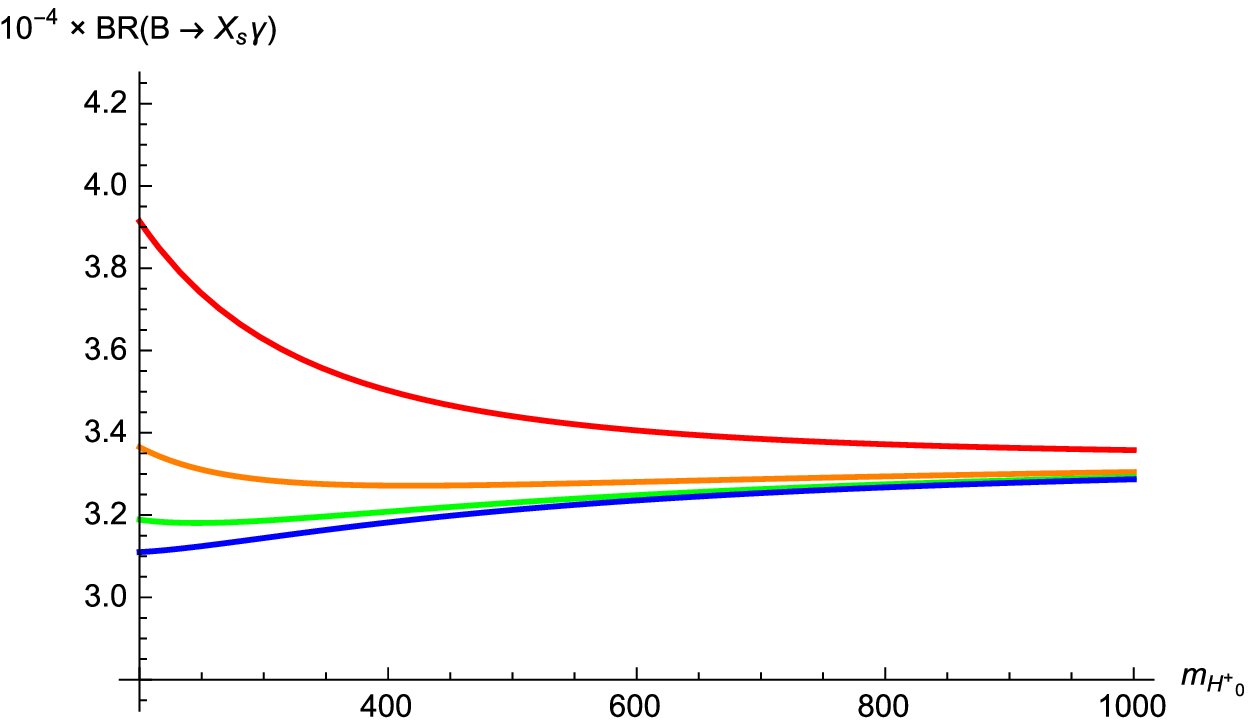} 
	\includegraphics[scale = 1]{Legend.eps} 
\caption{Branching ratio, $\mathcal{B}\left(B\longrightarrow X_s \gamma \right)$ shown for various LW scales. The upper (lower) left and right plots are calculated with the type-II (type-I) LW2HDM for $\tan\beta=1$ and $\tan\beta=2$ respectively. }
\label{fig:BXS}
\end{figure}

In order to compare to experimental data the calculation is carried out to NLO in QCD. The modifications to the amplitude are exactly the same as above LO example. The NLO amplitudes given in Ref. \cite{Ciuchini:1997xe} are those used in the numerical analysis. Numerical values used in the calculation are listed in the Appendix. Plots of the branching ratio are shown in Figure \ref{fig:BXS} for various LW scales for the type-I and II models.

\begin{figure}[H]
 \centering
 \begin{tabular}{cc}
 	\includegraphics[scale = .8]{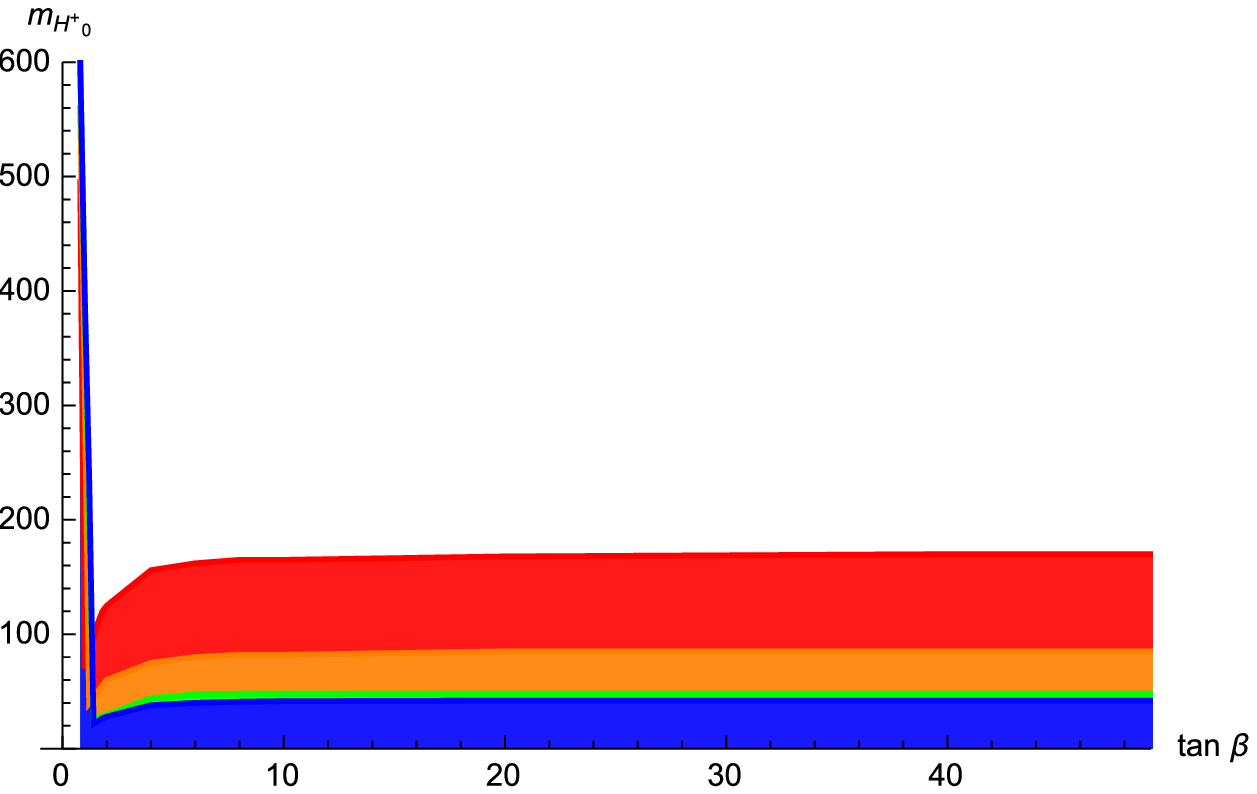} &
	\includegraphics[scale = 1]{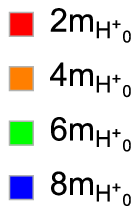} 
 \end{tabular}
\caption{Lower bounds on the mass of the charged Higgs, $m_{H^+_0}\,\, (GeV)$ from $B\longrightarrow X_S \gamma$ in the type-I LW2HDM at various Lee-Wick scales. }
\label{fig:bounds2}
\end{figure}

\begin{figure}[H]
 \centering
 	\includegraphics[scale = .5]{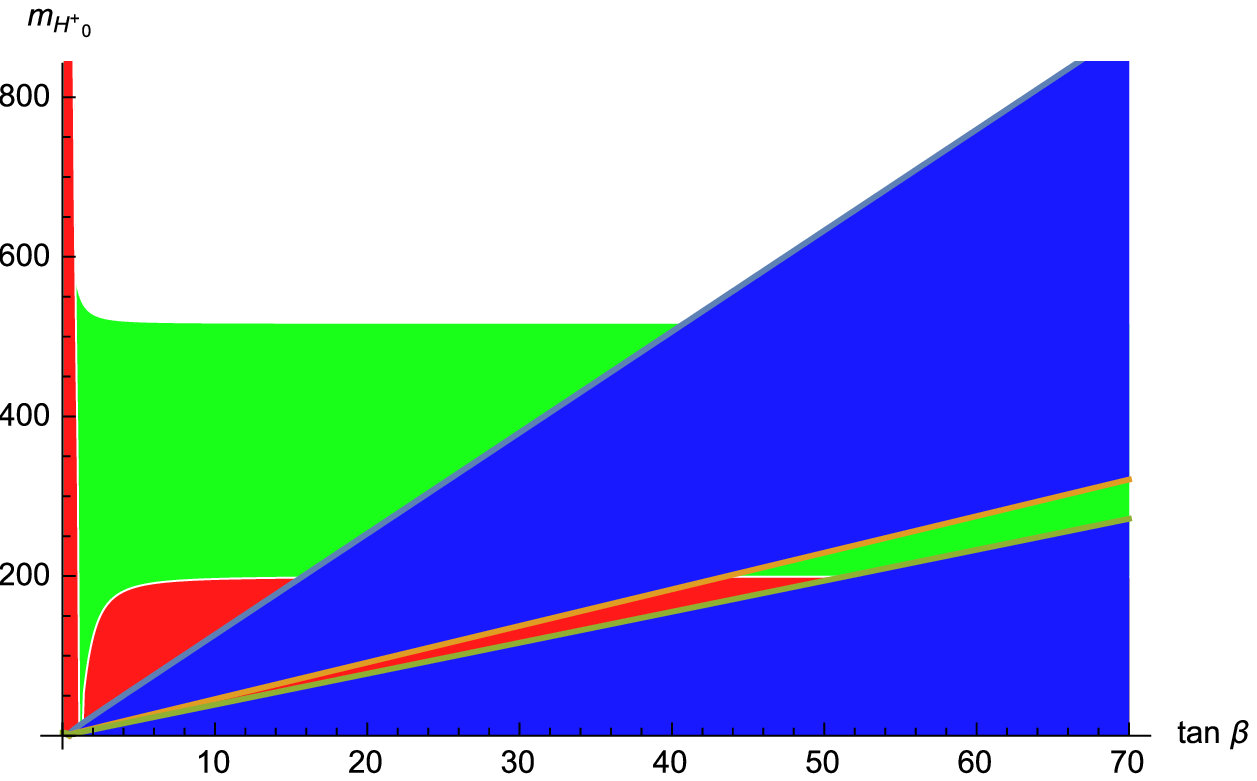}
	\includegraphics[scale = .5]{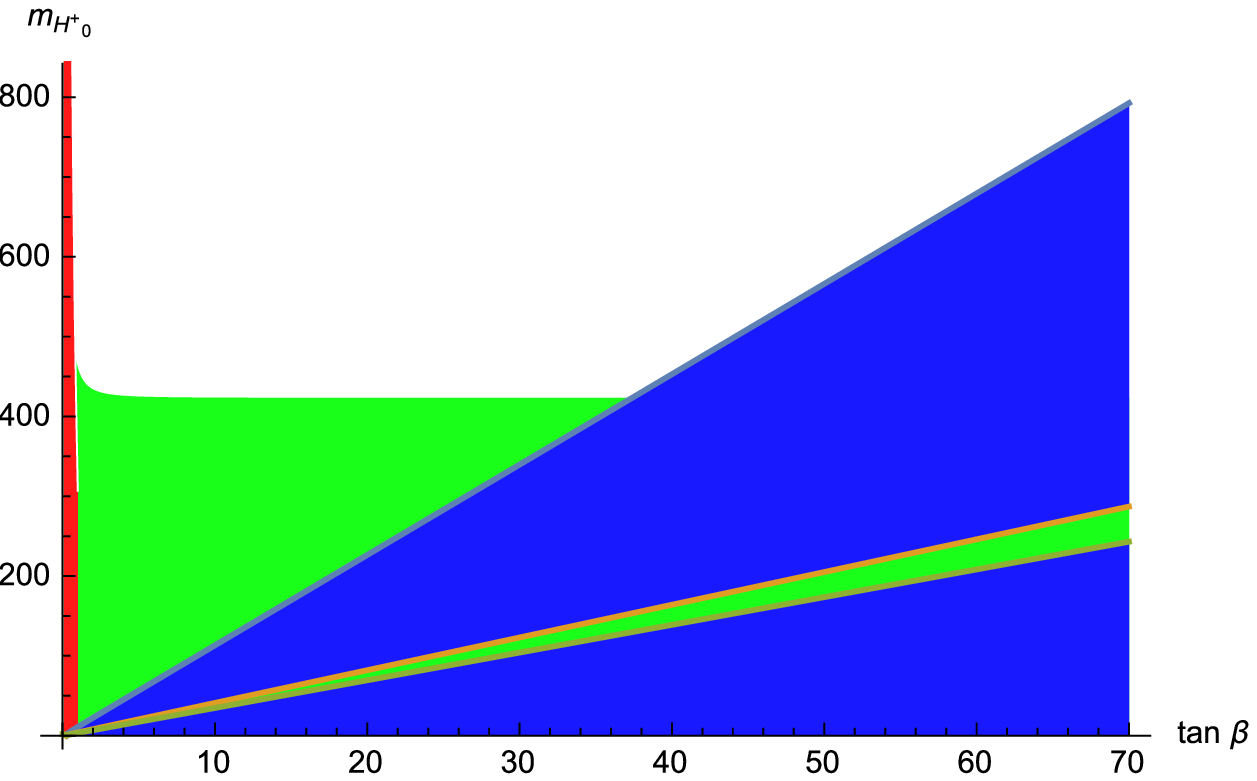} 
	\hspace{0mm}
	\includegraphics[scale = .5]{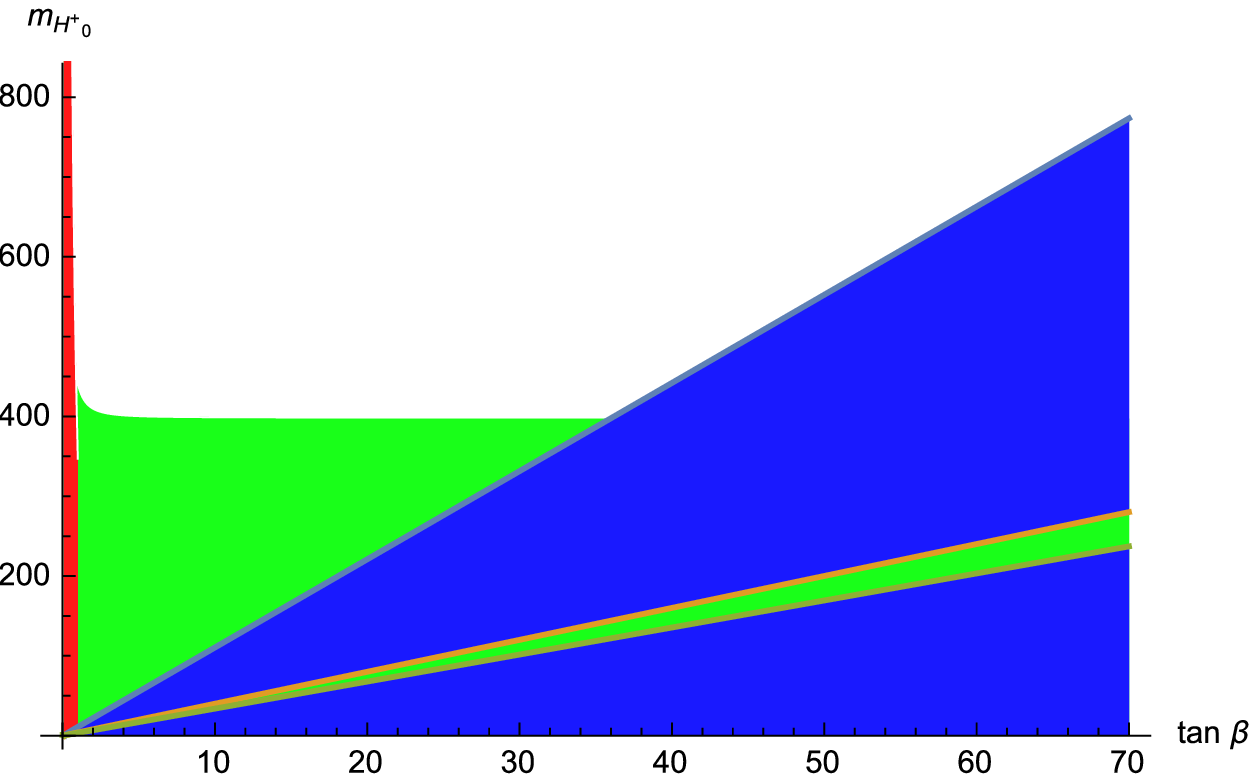} 
	\includegraphics[scale = 1]{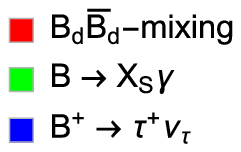} 
\caption{Lower bounds placed on the charged Higgs mass, $m_{H^+_0}\,\,(GeV)$ from B-physics constraints in the type-II LW2HDM. The plots are calculated with the Lee-Wick scales equal to $2\, m_{H_0^+}$ in the upper-left, $4\, m_{H_0^+}$ in the upper-right, and $8\, m_{H_0^+}$ on the bottom.}
\label{fig:bounds}
\end{figure}

The detected value for the branching ratio is $\mathcal{B}(B\longrightarrow X_s \gamma)=(3.52\pm0.23\pm0.09)\times 10^{-4}$ \cite{Barberio:2008fa}. As in the previous section, a $\chi^2$ text was used to establish lower bounds for the mass of the charged Higgs.  Plots of these bounds are shown in Figure \ref{fig:bounds} for the type-II model, and Figure \ref{fig:bounds2} for the type-I model. The bounds in the type-I model are qualitatively different in the LW2HDM as compared to the usual 2HDM result.  An asymptote occurs in the bounds of the model due to the couplings of the quarks to $\tilde{H}^\prime_0$ being independent of $\tan\, \beta$.  Below, plots of the lower bounds on the charged Higgs mass are shown for various Lee-Wick scales.

These bounds all apply to the charged Higgs masses.    Bounds on neutral Higgs masses are much weaker.   This is because all of the neutral scalars in the model couple in a flavor-diagonal way, and thus charged Higgs processes are the only ones that change flavor.   Bounds on flavor-changing processes are much stronger than those from flavor-conserving processes.  One potential low-energy effect is on the $\rho$-parameter, which is sensitive to the mass splitting within a isospin multiplet.  However, in this model the charged and neutral Lee-Wick scalars have very similar masses, thus this splitting is negligible.

\section{Results and Future Prospects}

From the B-physics results of the last section, the LW scale in the type-II model must exceed 800 GeV.  In the type-I model, the LW scale must exceed 400 GeV.  Is there a way to detect this at the LHC? 

Two possibilities for determining the validity of this theory exist.  The first involves changing the branching ratios of the $125$ GeV Higgs boson, and the other involves direct detection of LW states. 

Carone et al. \cite{Carone:2009nu} studied the effects of the LWSM on the decays of the Higgs boson and showed that current bounds are weak, with a lower bound of $255$ GeV on the LW scale.  They also noted that the bound will only become competitive with the B-decay bounds after 400 inverse femtobarns of integrated luminosity at the LHC.  Furthermore, the primary effect would be a slight increase in the $H\rightarrow \tau\tau$ branching ratio, making it unlikely that this would be interpreted as evidence for a LW sector.  Reaching the bound of $800$ GeV, as in the type-II version of the LW 2HDM above, would require an integrated luminosity in excess of $4000$ fb$^{-1}$, which is unlikely to be achieved in the next couple of decades.

Direct detection was discussed in detail by Figy and Zwicky \cite{Figy:2011yu}.  They wrote that the most likely discovery of the LW Higgs boson at the LHC would be if the mass was below the top pair production threshold (singular, since its the LW model, not 2HDM).  In addition, Figy and Zwicky noted that the negative width gives a dip-peak structure, instead of a peak-dip structure.  In this model the LW states are all above the top pair production threshold, making direct detection extremely difficult.  Detection of the LW states would require a substantially more energetic hadron collider or a multi-TeV linear collider.

Perhaps the best near-term hope for an indication of the LW2HDM model would be to discover the ``normal" particles of the 2HDM and study their decays.  The above Yukawa couplings differ from the conventional 2HDM. As a result, analysis of the Yukawa LW Higgs-coupled decays would provide evidence of the LW2HDM.

 \vskip 1.0cm
{\parindent = 0pt {\bf Acknowledgments}}
 \vskip 0.5cm
The authors would like to thank Chris Carone, Margarida Rebelo, Rui Santos and Josh Erlich for many useful discussions.   This work was supported by the NSF under Grants PHY-1068008 and PHY-1519644.
\newpage
\section{appendix}

The mass matrices in the Lee-Wick Two Higgs doublet model, using the basis \{$H_1,\, H_2,\, \tilde{H}_1,\, \tilde{H}_2$ \}:

\begin{center}
Charged Higgs 
\end{center}
\begingroup\makeatletter\def\f@size{8}\check@mathfonts
\be
\left(
\begin{array}{cccc}
 -\frac{v_2 \left(v_1 v_2 \left(\lambda _4+\lambda _5\right)-2 m_{12}^2\right)}{2 v_1} & \frac{1}{2} \left(v_1 v_2 \left(\lambda _4+\lambda _5\right)-2 m_{12}^2\right) & \frac{v_2 \left(v_1 v_2 \left(\lambda _4+\lambda _5\right)-2 m_{12}^2\right)}{2 v_1} & m_{12}^2-\frac{1}{2} v_1 v_2 \left(\lambda _4+\lambda _5\right) \\
 \frac{1}{2} \left(v_1 v_2 \left(\lambda _4+\lambda _5\right)-2 m_{12}^2\right) & -\frac{v_1 \left(v_1 v_2 \left(\lambda _4+\lambda _5\right)-2 m_{12}^2\right)}{2 v_2} & m_{12}^2-\frac{1}{2} v_1 v_2 \left(\lambda _4+\lambda _5\right) & \frac{v_1 \left(v_1 v_2 \left(\lambda _4+\lambda _5\right)-2 m_{12}^2\right)}{2 v_2} \\
 \frac{v_2 \left(v_1 v_2 \left(\lambda _4+\lambda _5\right)-2 m_{12}^2\right)}{2 v_1} & m_{12}^2-\frac{1}{2} v_1 v_2 \left(\lambda _4+\lambda _5\right) & -m_{\tilde{h}}^2-\frac{v_2 \left(v_1 v_2 \left(\lambda _4+\lambda _5\right)-2 m_{12}^2\right)}{2 v_1} & \frac{1}{2} \left(v_1 v_2 \left(\lambda _4+\lambda _5\right)-2 m_{12}^2\right) \\
 m_{12}^2-\frac{1}{2} v_1 v_2 \left(\lambda _4+\lambda _5\right) & \frac{v_1 \left(v_1 v_2 \left(\lambda _4+\lambda _5\right)-2 m_{12}^2\right)}{2 v_2} & \frac{1}{2} \left(v_1 v_2 \left(\lambda _4+\lambda _5\right)-2 m_{12}^2\right) & \frac{2 m_{12}^2 v_1-v_2 \left(2 m_{\tilde{h}}^2+v_1^2 \left(\lambda _4+\lambda _5\right)\right)}{2 v_2} \\
\end{array}
\right)
\ee
\endgroup

\vspace{2 mm}
\begin{center}
Neutral Pseudoscalar Higgs
\end{center}
\vspace{-3 mm}
\begingroup\makeatletter\def\f@size{8}\check@mathfonts
\be
\left(
\begin{array}{cccc}
 v_2 \left(\frac{m_{12}^2}{v_1}-v_2 \lambda _5\right) & v_1 v_2 \lambda _5-m_{12}^2 & v_2 \left(v_2 \lambda _5-\frac{m_{12}^2}{v_1}\right) & m_{12}^2-v_1 v_2 \lambda _5 \\
 v_1 v_2 \lambda _5-m_{12}^2 & v_1 \left(\frac{m_{12}^2}{v_2}-v_1 \lambda _5\right) & m_{12}^2-v_1 v_2 \lambda _5 & v_1 \left(v_1 \lambda _5-\frac{m_{12}^2}{v_2}\right) \\
 v_2 \left(v_2 \lambda _5-\frac{m_{12}^2}{v_1}\right) & m_{12}^2-v_1 v_2 \lambda _5 & v_2 \left(\frac{m_{12}^2}{v_1}-v_2 \lambda _5\right)-m_{\tilde{h}}^2 & v_1 v_2 \lambda _5-m_{12}^2 \\
 m_{12}^2-v_1 v_2 \lambda _5 & v_1 \left(v_1 \lambda _5-\frac{m_{12}^2}{v_2}\right) & v_1 v_2 \lambda _5-m_{12}^2 & v_1 \left(\frac{m_{12}^2}{v_2}-v_1 \lambda _5\right)-m_{\tilde{h}}^2 \\
\end{array}
\right)
\ee
\endgroup

\vspace{2 mm}
\begin{center}
Neutral Scalar Higgs
\end{center}
\vspace{-3 mm}
\begingroup\makeatletter\def\f@size{8}\check@mathfonts
\be
\left(
\begin{array}{cccc}
 \frac{v_2 m_{12}^2}{v_1}+v_1^2 \lambda _1 & v_1 v_2 \left(\lambda _3+\lambda _4+\lambda _5\right)-m_{12}^2 & -\frac{\lambda _1 v_1^3+m_{12}^2 v_2}{v_1} & m_{12}^2-v_1 v_2 \left(\lambda _3+\lambda _4+\lambda _5\right) \\
 v_1 v_2 \left(\lambda _3+\lambda _4+\lambda _5\right)-m_{12}^2 & \frac{v_1 m_{12}^2}{v_2}+v_2^2 \lambda _2 & m_{12}^2-v_1 v_2 \left(\lambda _3+\lambda _4+\lambda _5\right) & -\frac{\lambda _2 v_2^3+m_{12}^2 v_1}{v_2} \\
 -\frac{\lambda _1 v_1^3+m_{12}^2 v_2}{v_1} & m_{12}^2-v_1 v_2 \left(\lambda _3+\lambda _4+\lambda _5\right) & \frac{v_2 m_{12}^2}{v_1}-m_{\tilde{h}}^2+v_1^2 \lambda _1 & v_1 v_2 \left(\lambda _3+\lambda _4+\lambda _5\right)-m_{12}^2 \\
 m_{12}^2-v_1 v_2 \left(\lambda _3+\lambda _4+\lambda _5\right) & -\frac{\lambda _2 v_2^3+m_{12}^2 v_1}{v_2} & v_1 v_2 \left(\lambda _3+\lambda _4+\lambda _5\right)-m_{12}^2 & \frac{v_1 m_{12}^2}{v_2}-m_{\tilde{h}}^2+v_2^2 \lambda _2 \\
\end{array}
\right)
\ee
\endgroup

$$
v_1=v \cos (\beta )\,\,\,\,\,\,\,\,\,\,\,\,\,\,\,\,\,\,\,\,v_2=v \sin (\beta )\,\,\,\,\,\,\,\,\,\,\,\,\,\,\,\,\,\,\,\,m_{12}^2=\frac{1}{2} M_{12}^2 \sin (2 \beta );
$$

\begin{center}
Diagonalized pseudoscalar Higgs mass matrix
\end{center}
\vspace{-3 mm}
\begingroup\makeatletter\def\f@size{8}\check@mathfonts
\be
diag(0,\,m_{A^\pm_0}^2 ,\,-m_{\tilde{A}^{\prime\pm}_0}^2,\,-m_{\tilde{A}^\pm_0}^2 )=
\left(
\begin{array}{cccc}
 0 & 0 & 0 & 0 \\
 0 & -\frac{1}{2} m_{\tilde{h}}^2 \left(\sqrt{\frac{4 \lambda _5 v^2+m_{\tilde{h}}^2-4 M_{12}^2}{m_{\tilde{h}}^2}}-1\right) & 0 & 0 \\
 0 & 0 & -m_{\tilde{h}}^2 & 0 \\
 0 & 0 & 0 & -\frac{1}{2} m_{\tilde{h}}^2 \left(\sqrt{\frac{4 \lambda _5 v^2+m_{\tilde{h}}^2-4 M_{12}^2}{m_{\tilde{h}}^2}}+1\right) \\
\end{array}
\right)
\ee
\endgroup

\begin{center}
The diagonal elements the neutral scalar Higgs mass matrix
\end{center}
\begingroup\makeatletter\def\f@size{8}\check@mathfonts

\vspace{-3 mm}

$$
K=\sqrt{-2 M_{12}^2 v^2 \left(\lambda _{345} \sin ^2(2 \beta )+\cos (2 \beta ) \left(\lambda _1 \cos ^2(\beta )-\lambda _2 \sin ^2(\beta )\right)\right)+M_{12}^4+v^4 \left(\lambda _{345}^2 \sin ^2(2 \beta )+\left(\lambda _1 \cos ^2(\beta )-\lambda _2 \sin ^2(\beta )\right){}^2\right)}
$$
\be
m_{h_0}^2= -\frac{1}{2} m_{\tilde{h}}^2 \\
\left(\sqrt{\frac{m_{\tilde{h}}^2+2 K-2 M_{12}^2-2 v^2 \left(\lambda _2 \sin ^2(\beta )+\lambda _1 \cos ^2(\beta )\right)}{m_{\tilde{h}}^2}} -1\right)
\ee

\be
m_{H_0}^2=-\frac{1}{2} m_{\tilde{h}}^2 \\
\left(\sqrt{\frac{m_{\tilde{h}}^2-2 \left(K+M_{12}^2+v^2 \left(\lambda _2 \sin ^2(\beta )+\lambda _1 \cos ^2(\beta )\right)\right)}{m_{\tilde{h}}^2}}-1\right)
\ee

\be
-m_{\tilde{h}_0}^2=-\frac{1}{2} m_{\tilde{h}}^2 \left(\sqrt{\frac{m_{\tilde{h}}^2+2 K-2 M_{12}^2-2 v^2 \left(\lambda _2 \sin ^2(\beta )+\lambda _1 \cos ^2(\beta )\right)}{m_{\tilde{h}}^2}}+1\right)
\ee

\be
-m_{\tilde{H}_0}^2=-\frac{1}{2} m_{\tilde{h}}^2 \left(\sqrt{\frac{m_{\tilde{h}}^2-2 \left(K+M_{12}^2+v^2 \left(\lambda _2 \sin ^2(\beta )+\lambda _1 \cos ^2(\beta )\right)\right)}{m_{\tilde{h}}^2}}+1\right)
\ee
\endgroup

where $\lambda_{345} = \lambda_3+\lambda_4+\lambda_5$.   The scalar self-couplings are

\be
\lambda_1=\frac{\sec ^2(\beta ) \left(\sin ^2(\alpha ) m_{h_0}^2+\cos ^2(\alpha ) m_{H_0}^2\right)-M_{12}^2 \tan ^2(\beta )}{v^2}-\frac{\sec ^2(\beta ) \left(\sin ^2(\alpha ) m_{h_0}^4+\cos ^2(\alpha ) m_{H_0}^4\right)}{v^2 m_{\tilde{h}}^2}
\ee

\be
\lambda_2=\frac{\csc ^2(\beta ) \left(\cos ^2(\alpha ) m_{h_0}^2+\sin ^2(\alpha ) m_{H_0}^2\right)-M_{12}^2 \cot ^2(\beta )}{v^2}-\frac{\csc ^2(\beta ) \left(\cos ^2(\alpha ) m_{h_0}^4+\sin ^2(\alpha ) m_{H_0}^4\right)}{v^2 m_{\tilde{h}}^2}
\ee

\be
\lambda_{345}=\frac{\sin (\alpha ) \cos (\alpha ) \csc (\beta ) \sec (\beta ) \left(m_{h_0}^4-m_{H_0}^4\right)}{v^2 m_{\tilde{h}}^2}+\frac{\sin (2 \alpha ) \csc (2 \beta ) \left(m_{H_0}^2-m_{h_0}^2\right)+M_{12}^2}{v^2}
\ee

\be
\lambda_4=\frac{m_{A_0}^2-2 m_{H^\pm_0}^2+M_{12}^2}{v^2}-\frac{m_{A_0}^4-2 m_{H^\pm_0}^4}{v^2 m_{\tilde{h}}^2}
\ee

\be
\lambda_5=\frac{m_{A_0}^4}{v^2 m_{\tilde{h}}^2}+\frac{M_{12}^2-m_{A_0}^2}{v^2}
\ee

where $m_{h_0}, m_{H_0}, m_{A_0}, m_{H^\pm_0}$ and the two scalar masses, the pseudoscalar mass and the charged Higgs mass, respectively.
\vspace{5 mm}

Values used in calculations without explicit citation are from take from \cite{Amsler}.

\begin{table}[h]
\begin{tabular}{|c|c|}
\hline 
$m_{t}=171.2\pm2.1$GeV & $G_{F}=1.16637\times10^{-5}$ GeV$^{-2}$\tabularnewline
\hline 
$\bar{m}_{b}(\bar{m}_{b})=4.2_{-0.07}^{+0.17}$ GeV & $\alpha_{s}(m_{Z})=0.1176\pm$0.0020\tabularnewline
\hline 
$\bar{m}_{c}(\bar{m}_{c})=1.27_{-0.11}^{+0.07}$ GeV & $m_{B_{d}}=5279.53\pm0.33$ MeV\tabularnewline
\hline 
$m_{s}=104{}_{-34}^{+26}$ MeV & $f_{B}\sqrt{\hat{B}_{B_{d}}}=216\pm15$ MeV \cite{Gamiz:2009ku}\tabularnewline
\hline 
$m_{W}=80.398\pm0.025$ GeV & $\alpha_{em}^{-1}=137.03599967$\tabularnewline
\hline 
$m_{Z}=91.1876\pm0.021$ GeV & $\mathcal{B}(B\rightarrow X_{c}e\bar{\nu}_{e})=(10.74\pm0.16)\%$ \cite{Barberio:2008fa}\tabularnewline
\hline 
\end{tabular}
\end{table}

\end{document}